\documentclass[conference]{IEEEtran}
\IEEEoverridecommandlockouts
\usepackage{cite}
\usepackage{amsmath,amssymb,amsfonts}
\usepackage{algorithmic}
\RequirePackage[hidelinks]{hyperref}
\usepackage{graphicx}
\usepackage{textcomp}
\usepackage{xcolor}
\def\BibTeX{{\rm B\kern-.05em{\sc i\kern-.025em b}\kern-.08em
    T\kern-.1667em\lower.7ex\hbox{E}\kern-.125emX}}

\usepackage{floatrow}
\usepackage[label font=bf]{subfig}
\floatstyle{plaintop}
\restylefloat{table}

\newcommand*{\R}{\mathbb{R}}
\newcommand*{\X}{\mathcal{X}}
\newcommand*{\sign}{\text{sign}}

\begin{document}

\title{Equivariance-based self-supervised learning for audio signal recovery from clipped measurements
\thanks{VS's PhD is funded by ANR UNLIP. LJ's research is supported by the FRS-FNRS (QuadSense, T.0160.24).}
}

\author{
\IEEEauthorblockN{Victor Sechaud}
\IEEEauthorblockA{\textit{CNRS, ENS de Lyon} \\
\textit{Laboratoire de physique} \\
F-69007 Lyon, France \\
victor.sechaud@ens-lyon.fr}
\and
\IEEEauthorblockN{Laurent Jacques}
\IEEEauthorblockA{\textit{UCLouvain} \\
\textit{ICTEAM}\\
Louvain-la-Neuve, Belgium \\
laurent.jacques@uclouvain.be}
\and
\IEEEauthorblockN{Patrice Abry}
\IEEEauthorblockA{\textit{CNRS, ENS de Lyon} \\
\textit{Laboratoire de physique} \\
F-69007 Lyon, France \\
patrice.abry@ens-lyon.fr}
\and
\IEEEauthorblockN{Juli\'an Tachella}
\IEEEauthorblockA{\textit{CNRS, ENS de Lyon} \\
\textit{Laboratoire de physique} \\
F-69007 Lyon, France \\
julian.tachella@ens-lyon.fr}
}

\maketitle

\begin{abstract}
In numerous inverse problems, state-of-the-art solving strategies involve training neural networks from ground truth and associated measurement datasets that, however, may be expensive or impossible to collect.
Recently, self-supervised learning techniques have emerged, with  
the major advantage of no longer requiring ground truth data. 
Most theoretical and experimental results on self-supervised learning focus on linear inverse problems. 
The present work aims to study self-supervised learning for the non-linear inverse problem of recovering audio signals from clipped measurements.
An equivariance-based self-supervised loss is proposed and studied. 
Performance is assessed on simulated clipped measurements with controlled and varied levels of clipping, 
and further reported on standard real music signals. 
We show that the performance of the proposed equivariance-based self-supervised declipping strategy compares favorably to fully supervised learning while only requiring clipped measurements alone for training.
\end{abstract}


\section{Introduction}

Inverse problems appear in various and numerous scientific and engineering applications (such as tomography, MRI \cite{MRI_lustig2007sparse} or phase retrieval \cite{shechtman2015phase}). 
They consist in recovering a a signal $\boldsymbol{x} \in \X \subset \R^n$ from measurements $\boldsymbol{y} \in \mathcal{Y} \subset \R^m$, through a forward operator $A:\R^n\to\R^m$, possibly corrupted with noise $\epsilon \in \R^n$:
\begin{equation}
\boldsymbol{y} = A(\boldsymbol{x}) + \boldsymbol{\epsilon}.
\end{equation}
Inverse problems are often ill-posed, because the forward operator may be incomplete, \textit{i.e.}, $m<n$. 
Regularization strategies based on incorporating prior information have been widely studied and shown to have performance that strongly depends on the relevance of the chosen prior \cite{rudin1992nonlinear}. 
\begin{figure}
    \centering
    \includegraphics[width=1\textwidth]{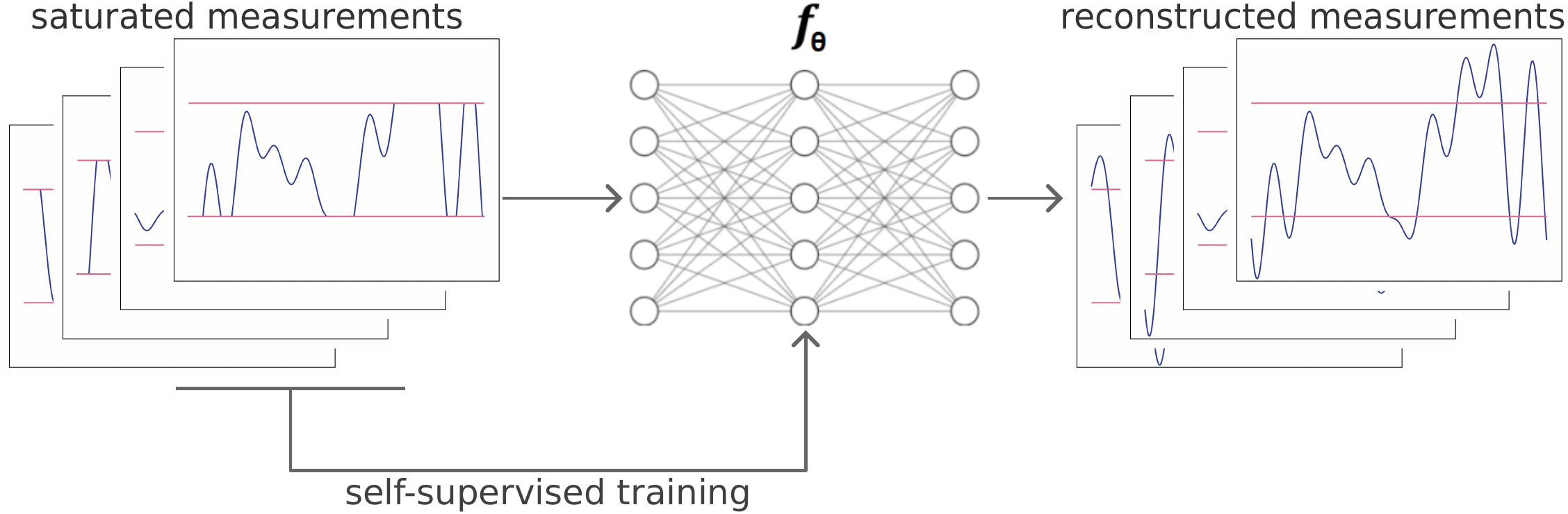}
    \caption{Learning to declip signals from measurement data alone. We propose a new self-supervised method that can learn to declip audio signals without ever seeing ground-truth reference signals by exploiting carefully chosen assumptions on the invariance of the reconstructed signal distribution.}
    \label{fig:enter-label}
\end{figure}
To overcome this limitation, most strategies are based on learning an inverse operator of $A$ on $\X$, $\ f_\theta: \boldsymbol{y} \mapsto \boldsymbol{x}$, from a training dataset containing ground truth signals and associated measurement $\left\{(\boldsymbol{x}_i,\boldsymbol{y}_i)\right\}^N_{i =1}$.
Standard approaches minimize the supervised mean square error (MSE): 
\begin{equation}
    \textstyle \min\limits_{\theta} \sum\limits^N_{i=1} \|f_\theta(\boldsymbol{y}_i)-\boldsymbol{x}_i\|^2 \label{MSE}
\end{equation}
where $f_\theta$ is generally modeled by a neural network.
Supervised learning however suffers from two main limitations: \emph{(i)} training sets can be difficult or impossible to obtain, \textit{e.g.}, in medical or space imaging \cite{lehtinen2018noise2noise}, and \emph{(ii)} even when ground-truth datasets are available, there might be a significant distribution shift between training (often synthetic) and testing (real) datasets.

To get over these two issues, self-supervised learning based on equivariance principles was proposed in \cite{chen2021equivariant} which only requires a dataset of measurements $\left\{\boldsymbol{y}_i\right\}^N_{i=1}$.
Their performance was theoretically assessed only for linear inverse problems \cite{Tachella2022SensingTF}. 

The present work contributes to extending equivariance-based self-supervised learning to non-linear inverse problems by addressing audio signal recovery from clipped measurements, as sketched in Fig.~\ref{fig:enter-label}. Declipping is a common nonlinear distortion, typically occurring with analog-to-digital (ADC) converters when the dynamic range of the original (analog) signal is too high.
\subsection{Related works}
\subsubsection{Audio declipping}
Several unsupervised techniques for audio declipping were proposed \cite{zavivska2020survey, boreal:134470, gaultier2021sparsity}, with many employing variational methods, incorporating priors such as sparsity in either the Fourier or time domain. 
Both $\ell_0$ and $\ell_1$ minimization approaches are used, often performed via algorithms like Douglas-Rachford or Chambolle-Pock. 
Additionally, learning-based methods such as dictionary learning were proven effective, with a dictionary trained from time windows of clipped signal $\boldsymbol{y}$. 
Other prior information inspired on human perception \cite{defraene2013declipping}, or leveraging the presence of data from multiple channels \cite{ozerov2016multichannel, gaultier2018cascade}, can be also used to improve the performance of these methods. 
The present work aims to avoid the reliance on such hand-crafted assumptions 
by learning directly from measurements alone.\\

\subsubsection{Self-supervised learning for inverse problems}
In the realm of unsupervised learning, several methods like noise2noise \cite{lehtinen2018noise2noise}, noise2void\cite{krull2019noise2void}, have gained prominence for handling noise in measurements. 
These methods primarily focus on mitigating noise artifacts but struggle with addressing incomplete information stemming from non-invertible operators $A$. 
An unsupervised learning method, referred to as equivariant imaging (EI), was proposed in \cite{chen2021equivariant} to solve linear inverse problems. 
It differs from noise2x methods by exploiting invariance properties, which may exist in natural images, to learn where $A$ does not provide information about the signal set $\X$. 
This is based on a natural intuition: 
the invariance of a signal set under a transformation implies that signals remain identifiable after the application of the transformation. 
Theoretical guarantees were provided in \cite{Tachella2022SensingTF} for linear forward operators only. 
A first attempt to study EI self-supervised learning for non-linear operators was conducted in \cite{tachella2023learning} in the one-bit compressed sensing framework, yet restricted to equivariance to translations and rotations. Here we show that equivariance to such geometrical transformations is not enough for learning from clipped measurements alone. 
\subsection{Goals, contributions, and outline} The goal of the present work is to devise and assess the performance of 
equivariance-based self-supervised learning to solve the nonlinear inverse problem of signal declipping, where amplitude invariance must be used, rather than translation or rotation invariance. 
To that end, Sec.~\ref{sec: unclipping} recalls the formulation of declipping as an inverse problem and discusses the \emph{naive} measurement consistency solutions and their limitations. 
Sec.~\ref{sec: EIunclipping} introduces scale invariance, discusses its use for equivariance learning, and designs the corresponding loss function. 
Sec.~\ref{sec: results} presents some numerical experiments.
First, synthetic simulated clipped measurements with controlled and varied levels of clipping are used, thus permitting quantifying the impact of the choice of hyperparameters on reconstruction performance.
Second, the application of the proposed equivariance-based self-supervised declipping strategy on a standard real sound dataset, shows that it performs comparably to fully supervised learning procedures.

\section{Unsupervised learning for declipping}
\label{sec: unclipping}

\subsection{Clipping forward model}
Clipping consists of applying a non-linear thresholding operator to the original data $x$: 
\begin{equation}
    \boldsymbol{y}=\eta(\boldsymbol{x}) \in \R^n, \label{clipping}
\end{equation} 
with $\eta$ the threshold operator of level $\mu >0$ (depending of the measurement device) defined for a scalar $c$:
\begin{equation*}
    \eta(c) = 
    \left\{\begin{array}{cl}
    \sign(c)\mu &\text{ if }  |c| \geq \mu \\
    c &\text{ otherwise }    
    \end{array}\right.
\end{equation*}
which is applied component-wise on vectors. 
\subsection{Naive measurement consistency}

In supervised learning, an MSE type loss,  similar to \eqref{MSE} is generally minimized. 
In unsupervised learning, the dataset consists of measurements only.
The training loss must thus be rewritten to avoid using the ground truth $\boldsymbol{x}$. 
The natural idea is to take the naive measurement consistency: 
\begin{equation}
    \mathcal{L}_{\textrm{NMC}}(\theta) = \sum\limits^N_{i=1} \left\|\boldsymbol{y}_i - \eta(f_\theta(\boldsymbol{y}_i))\right\|^2 \label{loss nmc}
\end{equation}
which enforces the reconstructed signals to be consistent with the observed measurements. 
However, it is impossible to learn from measurements consistency only because any function $f$ of the form 
\begin{equation}
    f(\boldsymbol{y})_j =
    \left\{\begin{array}{ccc}
        y_j & \text{ if } & |y_j| \leq \mu \\
         v(y_j) & \text{ if } & |y_j| =  \mu
    \end{array}\right.
\end{equation}
with $v$ being a function lying above the threshold level $\mu$ satisfies
\begin{equation}
    \eta(f(\boldsymbol{y})) = \boldsymbol{y}
\end{equation} 
and thus incurs zero training measurement loss.

\section{Equivariance-based self-supervised learning for declipping}
\label{sec: EIunclipping}

\subsection{Exploiting scale invariance}
To learn in the saturated part, we have to add another constraint that removes the limitation of measurement consistency alone. Here we assume the model $\X$ is amplitude invariant, \textit{i.e.} 
\begin{equation}
    \forall \boldsymbol{x} \in \X, \forall g \in \R_+, \ g\boldsymbol{x} \in \X. \label{amplitude invariant}
\end{equation} 
The intuition behind using this invariance is that each measurement can be seen as a signal through another threshold-type operator $\eta_g$:
\begin{equation}
    \eta(\boldsymbol{x})=\eta(gg^{-1}\boldsymbol{x}) := \eta_g (\boldsymbol{x'}). \label{familly of operators}
\end{equation}
with $\boldsymbol{x'} = g^{-1}\boldsymbol{x} \in \X$ and $\eta_g(\cdot) = \eta(g \cdot)$ by assumptions. 
Thus, we have a family of different operators with different saturation levels and different types of missing information. 
This invariance assumption differs from previous work on equivariant imaging \cite{chen2021equivariant, tachella2023learning} which relies on shift and/or rotations, as $\eta$ is equivariant to these geometrical group actions. Indeed, noting $T_g \in \R^{n\times n}$ as such transformations, we would have:
\begin{equation}
    \eta(\boldsymbol{x})=\eta(T_gT_{g^{-1}}\boldsymbol{x}) = T_g\eta(\boldsymbol{x'}),
\end{equation}
since $T_g$ are permutation matrices, and $\eta$ is applied elementwise. It results in a family of operators with the same saturation level (and thus do not provide additional information). We can also generalize \eqref{amplitude invariant} for other transformations beyond amplitude invariance, as long as \eqref{familly of operators} is respected, such as affine group transformation $\boldsymbol{x} \to a\boldsymbol{x} + \boldsymbol{b}$, with $g = (a,\boldsymbol{b}) \in \R_+ \times \R^n$.

In practice, to exploit this invariance property we remark that since $g\boldsymbol{x} \in \X$, $f$ has to be able to reconstruct $\boldsymbol{x}$ and $g\boldsymbol{x}$ for all $\boldsymbol{x}$ in $\X$, \textit{i.e.}
 \begin{equation}
     f\left(\eta(g\boldsymbol{x})\right) = g\boldsymbol{x} \text{ and }f(\eta(\boldsymbol{x})) = \boldsymbol{x}
 \end{equation}
 thus
 \begin{equation}
     f(\eta(g\boldsymbol{x})) = g(f(\eta(\boldsymbol{x})).
 \end{equation}
In other words, the composition of $f$ with $\eta$ should be equivariant for the multiplicative group $(\R_+, \cdot)$.
\subsection{Proposed loss}
\begin{figure}
    \centering
    \includegraphics[width=1\textwidth]{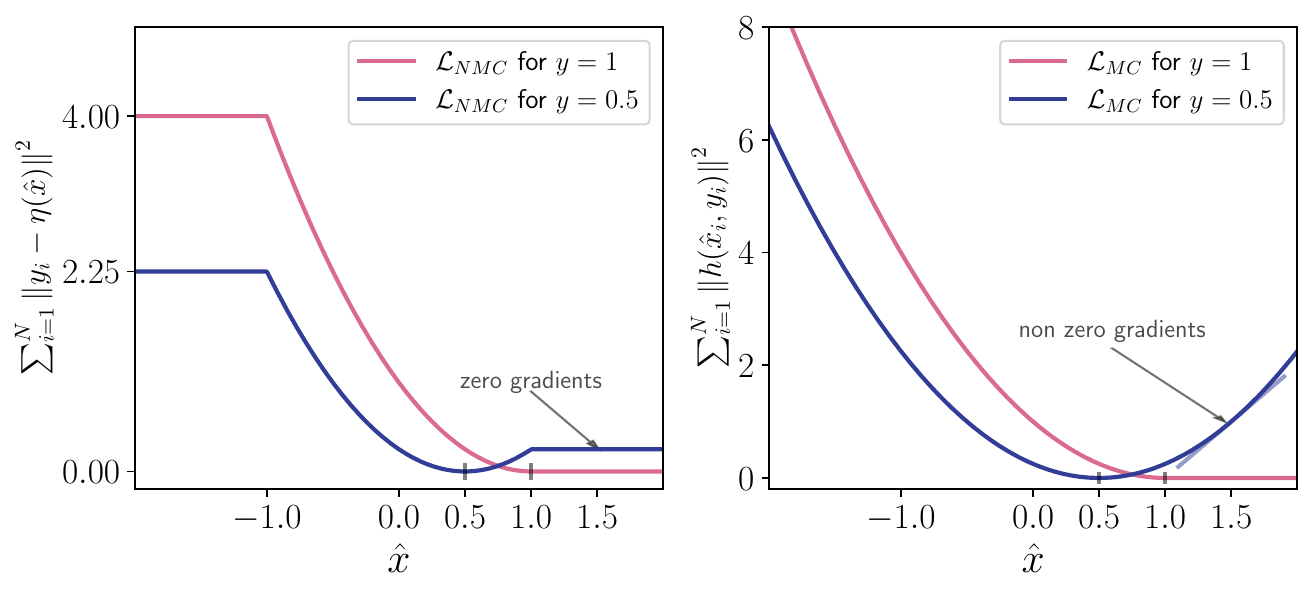}
    \caption{Comparison of $\mathcal{L}_{\textrm{NMC}}$ and $\mathcal{L}_{\textrm{MC}}$ with $\mu= 1$. On the left: Curve of the $\mathcal{L}_{\textrm{NMC}}$ in one dimension for two examples in both cases saturated and not. On the right: Curve of the new $\mathcal{L}_{\textrm{MC}}$ for the same two examples.}
    \label{fig: nmc vs mc}
\end{figure}

Our approach first considers a loss function whose minimization enforces the equivariance of the reconstruction function to the multiplicative group. We define it as:
\begin{equation}
    \mathcal{L}_{\textrm{EI}}(\theta) =  \sum\limits^N_{i=1}\mathbb{E}_{g\sim p_g}\left[  \left\| gf_\theta\left(\boldsymbol{y}_i\right) - f_\theta\left(\eta\left(gf_\theta\left(\boldsymbol{y}_i\right)\right)\right)\right\|^2\right] \label{loss ei}
\end{equation} 
where the expectation is computed with respect to a certain distribution $p_g$ over the multiplicative group. The choice of this distribution is further discussed in Sec.~\ref{sec: results}.
Second, our loss function should also include a second term that promotes measurement consistency. While the simple loss introduced in \eqref{loss nmc} fulfills this role, its gradient vanishes where the signal is not saturated ($|y_j| < \mu$). Indeed, in \eqref{loss nmc}, as $\eta$ has zero gradients beyond the threshold levels, we cannot hope to have the direction that minimizes the loss (by gradient descent) whereas, if the loss function is adequate, its gradient should point towards $y$. We therefore choose to use the MSE where the signal is not saturated, and to alter it otherwise according to the following loss: 
\begin{equation}
    \mathcal{L}_{\textrm{MC}}(\theta)= \sum\limits^N_{i=1} \left\|h(f_\theta(\boldsymbol{y}_i), \boldsymbol{y}_i)\right\|^2.
\end{equation}
with $h$ defined as
\begin{equation*}
h(\boldsymbol{u}, \boldsymbol{v})_j = 
    \left\{\begin{array}{ccl}
        v_j-u_j  &  \text{ if } & |v_j| < \mu \text{ or } \sign(v_j) u_j < \mu \\
        0  & \text{ if } & |v_j|= \mu \text{ and } \sign(v_j) u_j \geq \mu.
    \end{array}\right.
\end{equation*}
See Fig.~\ref{fig: nmc vs mc} for an example with a single signal in one dimension.

Adding both losses together results in a new loss
\begin{equation}
   \mathcal{L}(\theta) = \mathcal{L}_{\textrm{MC}}(\theta) + \mathcal{L}_{\textrm{EI}}(\theta)
\end{equation}
which enforces both consistency with the first term and equivariance with the second.

\subsection{Bias free reconstruction}
\begin{figure}[htbp]
    \centering
     \includegraphics[width=1\textwidth]{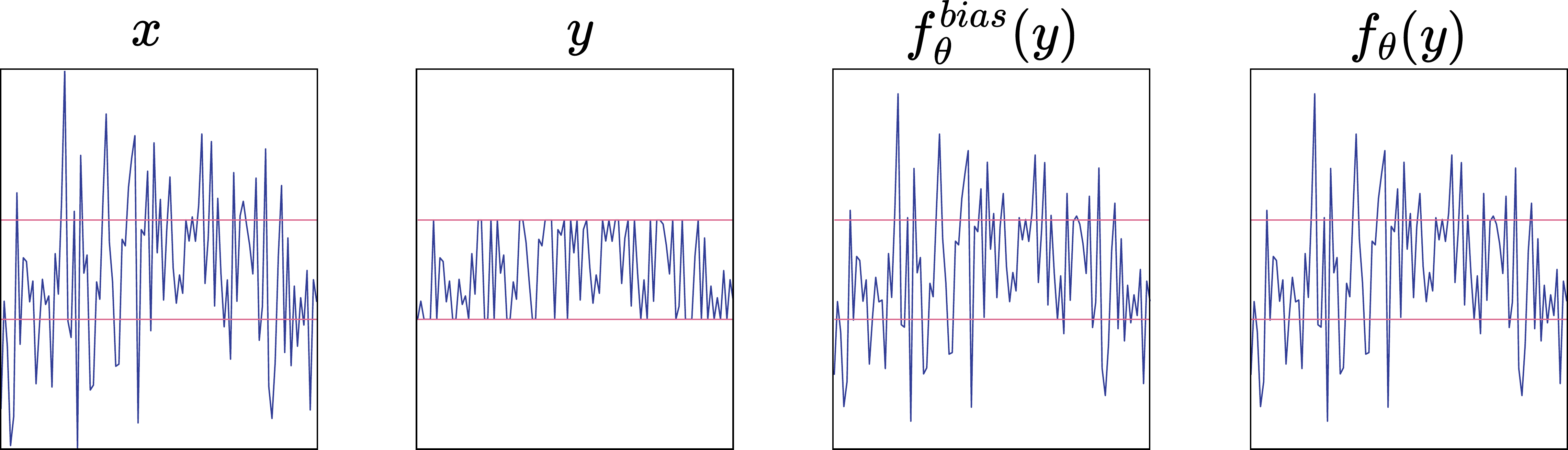}
     \vspace*{0mm}
    \vspace{-3mm}
    
     \includegraphics[width=1\textwidth]{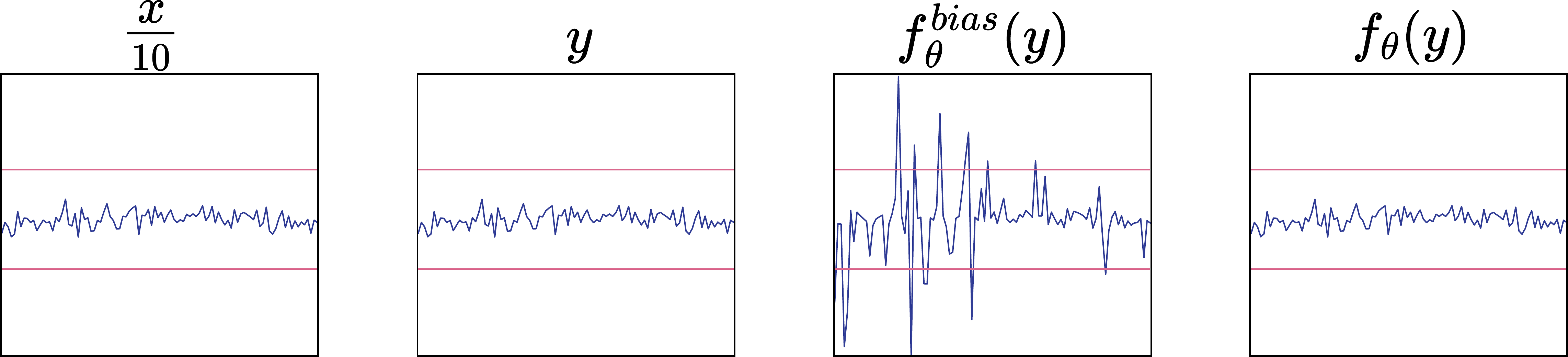}
    \caption{A toy example where a network with learnable biases is not suitable for a scale-invariant signal set. From left to right: Top row: a signal, the associated measurement, and reconstruction; Bottom row: the same signal divided by 10 (still contained in $\X$ by assumption) with the associated measurement and reconstruction.}
    \label{fig: biased network}
\end{figure}

We choose a bias-free version of ReLU neural network, to respect the scale-invariant assumption as proposed in \cite{mohan2019robust}. Indeed a feedforward neural network with ReLU activation functions and without bias is a homogeneous function:
\begin{equation}
    f(\alpha y) = \alpha f(y) \quad \forall \alpha \in \R_+.
\end{equation}
While this architectural choice does not render the system $f\circ\eta$ amplitude equivariant, it can still improve the generalization properties of $f$, since this is a desirable property for the non-saturated parts of $y$ as it is illustrated in Fig.~\ref{fig: biased network}.

\subsection{Masking}
The learned reconstruction network can often slightly modify the unsaturated part of the signal. To remove this undesired effect, at test time, we compute a mask that modifies only saturated measurements, \textit{i.e.} 
\begin{equation}
    \hat{x}_j = (1-b_j)y_j + b_j f_\theta(y)_j.
    \label{blending}
\end{equation}
\begin{equation*}
    \text{with } b_j = \dfrac{\max(0, |y_j| -\tau\mu)}{1-\tau\mu}.    
\end{equation*}
for all $j=1,\dots, n$, where $\tau$ is an hyper parameter set to $0.95$ in all the experiments.

Another way to use information from unsaturated parts is to feed it into the neural network. For this purpose, we add a mask of the unsaturated part as a channel to the signal, before being fed into the neural network. In theory, we expect that the neural network itself learns to calculate this mask, but experimentally we have noticed that explicitly providing the mask as an input improves the results.

\begin{figure}
    \centering
    \includegraphics[width=1\textwidth]{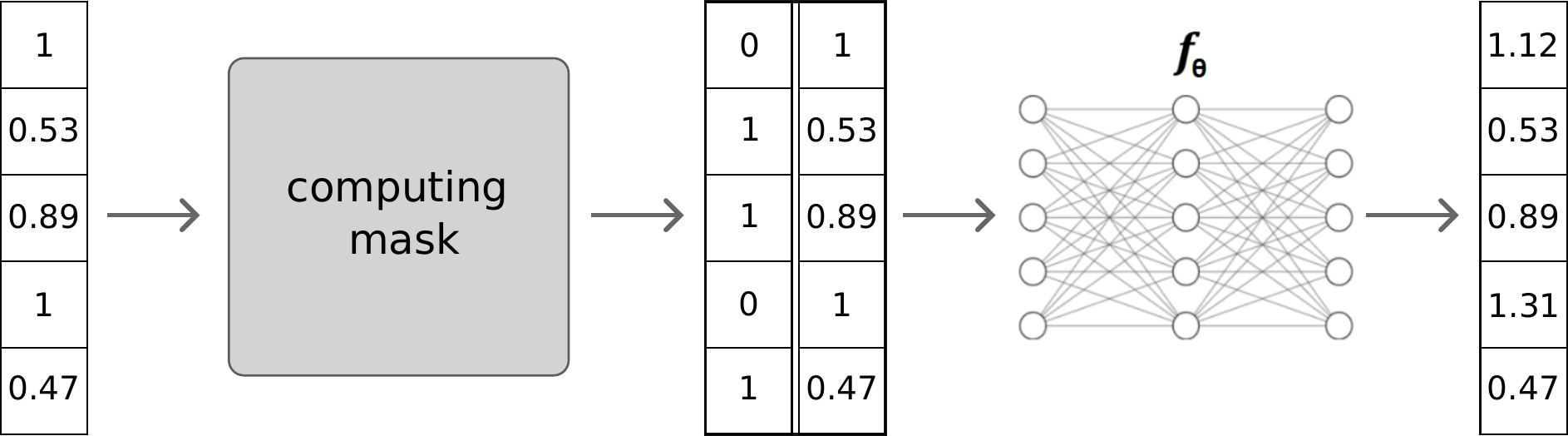}
    \caption{Diagram illustrating the used model. A mask is computed from the measurement and both the mask and measurements are fed to the network.}
\end{figure}

\section{Experiments} 
\label{sec: results}
In this section, we present a series of experiments that evaluate the performance of the proposed method. Two datasets are used, one synthetic which allows us to handle the proportion of saturated samples and the size of the model, and the other corresponding to real sounds. The performance is evaluated using the Signal Distortion Ratio (SDR) between the ground-truth $\boldsymbol{x}$ and the reconstruction $\hat{\boldsymbol{x}}$ \cite{zavivska2020survey}:
\begin{equation}
    \textrm{SDR}(\boldsymbol{x},\hat{\boldsymbol{x}}) = 20\log_{10}\left( \dfrac{\|\boldsymbol{x}\|_2}{\|\boldsymbol{x} - \hat{\boldsymbol{x}}\|_2} \right).
\end{equation}

\subsection{Synthetic dataset}
\begin{figure}
    \centering
    \includegraphics[width=1\textwidth]{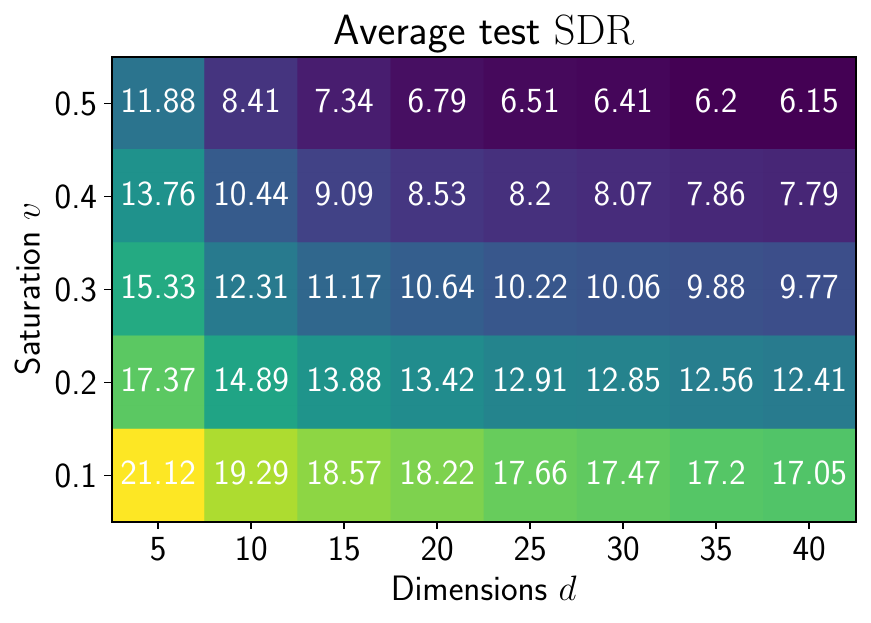}
    \caption{Average reconstruction performance as a function of saturated part $v$ and model dimension $d$. The value indicated corresponds to the mean $\textrm{SDR}$ over the test dataset.}
    \label{fig: tab sdr}
\end{figure}

We start with experiments on a synthetic dataset constructed as follows: first, we generate a random universal subspace of $\R^{100}$ with dimension $d \in \left\{ 1,\dots, 100 \right\}$ from basis vectors which each coordinate drawn from a standard normal distribution. 
We then generate $n = 1000$ vectors $\left\{\boldsymbol{x}\right\}^N_{i=1}$ on this subspace, which we rescale by a factor $q_i \in \R_+$ chosen so that a proportion $v \in (0, 1)$ of the resulting measurements are clipped (with the threshold level $\mu$ set to 1). 
We evaluate the recovery performance depending on the parameters $d$ and $v$.
Fig.~\ref{fig: tab sdr} demonstrates the network's ability to reconstruct unsaturated signals, as a function of these two parameters. The test performance diminishes with a larger subspace dimension or an increased number of saturated samples. Indeed signals belonging to a higher dimensional subspace are harder to reconstruct like studying in the context of sparsity \cite{foucart2017sparse}. 

For this experiment, the distribution $p_g$ referred in \eqref{loss ei} is the uniform distribution on $[0.5, 1.5]$. We use a fully connected multilayer perceptron (MLP) with skip-connection for this experiment to remove any inductive biases associated with convolutional networks \cite{ulyanov2018deep}.

\subsection{Real dataset}
Here we evaluate the method on a dataset of real sounds. We use the GTZAN dataset, which is composed of a collection of 10 genres with 100 audio files each, all having a length of 30 seconds with a sample rate of 22 050~Hz. We focus here on classical music but similar results could be obtained for each genre or by mixing them. Sounds are split into 30 1-second signals and they are then saturated with a threshold level $\mu = 0.1$. Finally, measurements $\boldsymbol{y}$ that don't have any saturated entry are removed. This results in a training dataset of 576 sounds and a testing dataset of size 63. For these experiments, we choose a standard UNet architecture which is well adapted for the time correlation of natural sound signals; and the uniform distribution on $[0.1, 2]$ for $p_g$. Tab.~\ref{tab: SDR no mask} shows that the proposed self-supervised method obtains a performance on par with the fully supervised method. Tab.~\ref{tab: SDR with mask} shows that using the mask can help to improve performance by approximately 0.5 dB.

\begin{table}
\caption{Reconstruction performance with blending for both supervised and unsupervised methods without using a mask in network input. SDR is averaged over the test dataset.}
\centering
\begin{tabular}{c|c}
        Methods & SDR  \\ \hline
        Identity & $10.39 \pm 3.77$ \\ 
        Supervised & $16.45 \pm 3.83$ \\ 
        Unsupervised & $15.28 \pm 3.81$ \\ \hline
\end{tabular}
\label{tab: SDR no mask}
\end{table}

\begin{table}
\centering
\caption{Reconstruction performance for both mask used and not. SDR is averaged over all test dataset.}
\begin{tabular}{c|c}
        Methods & SDR  \\ \hline
        Identity & $10.39 \pm 3.77$ \\ 
        Supervised using mask & $16.43 \pm 3.81$ \\ 
        Unsupervised without mask & $15.28 \pm 3.81$ \\ 
        Unsupervised using mask & $15.8 \pm 4.26$ \\ \hline
\end{tabular}
\label{tab: SDR with mask}
\end{table}

To illustrate that it is possible to have a supervised training dataset that does not transfer well to the test set, we perform the following experiment: we create a supervised training dataset with music only and a test dataset with music and voices \cite{vryzas2018speech, vryzas2018subjective} containing only measurements (no ground truth measurements). The supervised method (trained using \eqref{MSE}) learns from the training dataset and the self-supervised on the train and test datasets because it does not need ground truth, they are then both evaluated on the test dataset. The result presented in Tab.~\ref{tab: robust self-sup} shows that the self-supervised method is more robust when the training dataset differs from the test set. 
\begin{table}
\caption{Comparison between supervised and self-supervised method, the first train on music dataset with ground-truth, the second music and voices dataset without ground-truth. both results are evaluated using the voices dataset.}
\centering
\begin{tabular}{c|c}
        Methods & SDR  \\ \hline
        Identity & $6.54 \pm 2.34$ \\ 
        Supervised & $10.94 \pm 2.00$ \\ 
        Unsupervised & $11.92 \pm 2.46$ \\ \hline
\end{tabular}
\label{tab: robust self-sup}
\end{table}

\subsection{Influence of parameter choices}
\begin{figure}
    \centering
    \includegraphics[width=.8\textwidth]{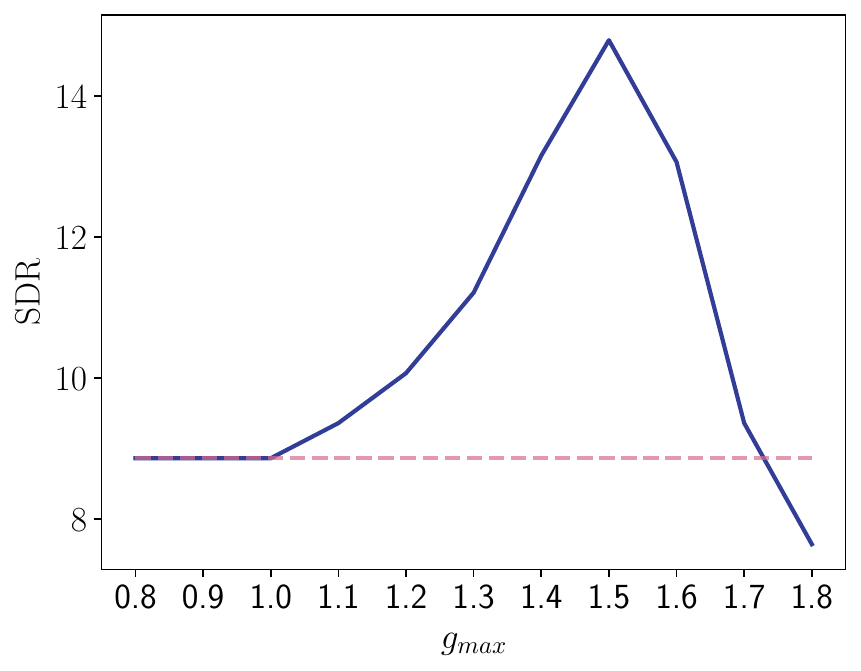}
    \caption{Average reconstruction performance as a function of $g_{\textrm{max}}$. The dashed line corresponds to the mean SDR of the test synthetic dataset.}
    \label{fig: sdr_function_gmax}
\end{figure}

The loss function \eqref{loss ei} depends on a distribution $p_g$ which in theory could be any unbounded distribution over $\R_{+}$, but in practice, signals have a bounded amplitude, and distribution over a bounded space $[g_{\textrm{min}},g_{\textrm{max}}]$ (not necessary uniform) may therefore suffice. Parameters have been chosen empirically: if $g_{\textrm{max}}$ is too small, the network will not learn; if it is too large, learning becomes unstable, making it difficult to converge. 
Fig~\ref{fig: sdr_function_gmax} shows how the reconstruction behaves as a function of the value of $g_{\textrm{max}}$ for the synthetic dataset with $v =0.3$ and $d=10$. The best performance in this case is obtained with $g_{\textrm{max}} = 1.5$. 

\section{Conclusion}
In this study, we propose an unsupervised method that uses scale invariance to address the nonlinear saturation problem. Experimental results indicate that this method can perform on par with the standard supervised approach, but on average, it does not surpass it, since it uses strictly less information for training. The high variance observed in both methods suggests significant variability in the results.
Looking ahead, there are several possible directions for future research. Firstly, providing theoretical guarantees for the proposed method could enhance its applicability and understanding. Additionally, integrating additional prior knowledge, such as multi-channel or psychoacoustics, may improve the method's robustness and performance in some scenarios.
As the saturation problem is not specific to sounds, experiments on other databases such as images can be envisaged.  Finally, we envisage extending our method to other non-linear problems beyond signal saturation, such as phase retrieval.
\bibliographystyle{ieeetr}
\bibliography{biblio}

\begin{thebibliography}{10}

\bibitem{MRI_lustig2007sparse}
M.~Lustig, D.~Donoho, and J.~M. Pauly, ``Sparse mri: The application of
  compressed sensing for rapid mr imaging,'' {\em Magnetic Resonance in
  Medicine: An Official Journal of the International Society for Magnetic
  Resonance in Medicine}, vol.~58, no.~6, pp.~1182--1195, 2007.

\bibitem{shechtman2015phase}
Y.~Shechtman, Y.~C. Eldar, O.~Cohen, H.~N. Chapman, J.~Miao, and M.~Segev,
  ``Phase retrieval with application to optical imaging: a contemporary
  overview,'' {\em IEEE signal processing magazine}, vol.~32, no.~3,
  pp.~87--109, 2015.

\bibitem{rudin1992nonlinear}
L.~I. Rudin, S.~Osher, and E.~Fatemi, ``Nonlinear total variation based noise
  removal algorithms,'' {\em Physica D: nonlinear phenomena}, vol.~60, no.~1-4,
  pp.~259--268, 1992.

\bibitem{lehtinen2018noise2noise}
J.~Lehtinen, J.~Munkberg, J.~Hasselgren, S.~Laine, T.~Karras, M.~Aittala, and
  T.~Aila, ``Noise2noise: Learning image restoration without clean data,'' {\em
  arXiv preprint arXiv:1803.04189}, 2018.

\bibitem{chen2021equivariant}
D.~Chen, J.~Tachella, and M.~E. Davies, ``Equivariant imaging: Learning beyond
  the range space,'' pp.~4379--4388, 2021.

\bibitem{Tachella2022SensingTF}
J.~Tachella, D.~Chen, and M.~E. Davies, ``Sensing theorems for unsupervised
  learning in linear inverse problems,'' {\em J. Mach. Learn. Res.}, vol.~24,
  pp.~39:1--39:45, 2022.

\bibitem{zavivska2020survey}
P.~Z{\'a}vi{\v{s}}ka, P.~Rajmic, A.~Ozerov, and L.~Rencker, ``A survey and an
  extensive evaluation of popular audio declipping methods,'' {\em IEEE Journal
  of Selected Topics in Signal Processing}, vol.~15, no.~1, pp.~5--24, 2020.

\bibitem{boreal:134470}
S.~Kitic, L.~Jacques, N.~Madhu, M.~P. Hopwood, A.~Spriet, and
  C.~De~Vleeschouwer, ``Consistent iterative hard thresholding for signal
  declipping,'' in {\em Acoustics, Speech and Signal Processing (ICASSP), 2013
  IEEE International Conference on}, IEEE, 2013.

\bibitem{gaultier2021sparsity}
C.~Gaultier, S.~Kiti{\'c}, R.~Gribonval, and N.~Bertin, ``Sparsity-based audio
  declipping methods: Selected overview, new algorithms, and large-scale
  evaluation,'' {\em IEEE/ACM Transactions on Audio, Speech, and Language
  Processing}, vol.~29, pp.~1174--1187, 2021.

\bibitem{defraene2013declipping}
B.~Defraene, N.~Mansour, S.~De~Hertogh, T.~Van~Waterschoot, M.~Diehl, and
  M.~Moonen, ``Declipping of audio signals using perceptual compressed
  sensing,'' {\em IEEE Transactions on Audio, Speech, and Language Processing},
  vol.~21, no.~12, pp.~2627--2637, 2013.

\bibitem{ozerov2016multichannel}
A.~Ozerov, {\c{C}}.~Bilen, and P.~P{\'e}rez, ``Multichannel audio declipping,''
  in {\em 2016 IEEE International Conference on Acoustics, Speech and Signal
  Processing (ICASSP)}, pp.~659--663, IEEE, 2016.

\bibitem{gaultier2018cascade}
C.~Gaultier, N.~Bertin, and R.~Gribonval, ``Cascade: Channel-aware structured
  cosparse audio declipper,'' in {\em 2018 IEEE International Conference on
  Acoustics, Speech and Signal Processing (ICASSP)}, pp.~571--575, IEEE, 2018.

\bibitem{krull2019noise2void}
A.~Krull, T.-O. Buchholz, and F.~Jug, ``Noise2void-learning denoising from
  single noisy images,'' in {\em Proceedings of the IEEE/CVF conference on
  computer vision and pattern recognition}, pp.~2129--2137, 2019.

\bibitem{tachella2023learning}
J.~Tachella and L.~Jacques, ``Learning to reconstruct signals from binary
  measurements alone,'' {\em Transactions on Machine Learning Research}, 2023.
\newblock Featured Certification.

\bibitem{mohan2019robust}
S.~Mohan, Z.~Kadkhodaie, E.~P. Simoncelli, and C.~Fernandez-Granda, ``Robust
  and interpretable blind image denoising via bias-free convolutional neural
  networks,'' {\em arXiv preprint arXiv:1906.05478}, 2019.

\bibitem{foucart2017sparse}
S.~Foucart and T.~Needham, ``Sparse recovery from saturated measurements,''
  {\em Information and Inference: A Journal of the IMA}, vol.~6, no.~2,
  pp.~196--212, 2017.

\bibitem{ulyanov2018deep}
D.~Ulyanov, A.~Vedaldi, and V.~Lempitsky, ``Deep image prior,'' in {\em
  Proceedings of the IEEE conference on computer vision and pattern
  recognition}, pp.~9446--9454, 2018.

\bibitem{vryzas2018speech}
N.~Vryzas, R.~Kotsakis, A.~Liatsou, C.~A. Dimoulas, and G.~Kalliris, ``Speech
  emotion recognition for performance interaction,'' {\em Journal of the Audio
  Engineering Society}, vol.~66, no.~6, pp.~457--467, 2018.

\bibitem{vryzas2018subjective}
N.~Vryzas, M.~Matsiola, R.~Kotsakis, C.~Dimoulas, and G.~Kalliris, ``Subjective
  evaluation of a speech emotion recognition interaction framework,'' in {\em
  Proceedings of the Audio Mostly 2018 on Sound in Immersion and Emotion},
  pp.~1--7, 2018.

\end{thebibliography}

\end{document}